# Millivolt modulation of a plasmonic metasurface via ionic conductance

*Krishnan Thyagarajan, [1,2], Ruzan Sokhoyan [1], Leonardo Z. Zornberg [1] and Harry A. Atwater [1,2]*

[1] Thomas J. Watson Laboratory of Applied Physics, California Institute of Technology, Pasadena, California 91125, USA

[2] Kavli Nanoscience Institute, California Institute of Technology, Pasadena, California 91125, USA

**ABSTRACT:** We report here and experimentally demonstrate an actively controlled gate-tunable plasmonic metasurface operating in the visible region of the electromagnetic spectrum, where – strikingly – the operating voltages for reflectance modulation are much less than 1V. The electrically tunable metasurface consists of inverse dolmen structures (iDolmen) patterned on silver and chromium on a quartz substrate and subsequently covered with a 5 nm thin layer of $Al_2O_3$ followed by a 110 nm indium tin oxide (ITO) layer, which acts as a transparent electrode. Our designed structures show up to 78% change in reflection upon applying small voltages (<1V). We explain this behaviour via ion conductance of silver through $Al_2O_3$ and ITO, leading to active resistive switching. Interesting complementary effects such as decreased reflection in the same structures over a broadband of wavelengths is also seen on reversing the applied bias. The results provide an insight into the use of the resistive switching for electrical control over light-matter interaction in plasmonic metasurfaces.

**ARTICLE:**

Controlling the configuration of nanostructures in the mesoscopic domain, permits targeted light-matter interaction allowing the creation of systems including optical modulators, switches, phase plates and wave-front shaping components. Metasurfaces are two dimensional analogues of metamaterials, and are comprised of nanostructures on a surface that modifies the way incident light interacts with the surface. Many examples of such metasurfaces have been explored in the past and have given rise to the field of flat optics, where conventional optical functionalities are retained while shrinking the macroscopic optical component itself, into a two dimensional surface. Such a modification of the surface has been demonstrated in earlier work to effectively control various facets of light-matter interaction including the phase,[1] polarization,[2] reflection and transmission amplitude,[3] the shaping of a complex wavefront[4] etc. However such metasurfaces are passive in nature, implying that their response is fixed once they are fabricated. Another class of metasurfaces includes those that are active in nature. Such systems can be made to dynamically change their response to the incident light by using an external mechanism. Many different mechanisms have been exploited to make such active metasurfaces. These include the use of liquid crystals,[5] the modulation of fermi levels in graphene by the application of an external voltage bias,[6] the change in the phase of a material upon the application or removal of heat,[7] the field effect in materials such as transparent conducting oxides which exhibit large changes in their complex refractive indices upon the application of a bias[8] etc. It is hoped that a few of these mechanisms may be used for the active optoelectronic control of elements on a chip since the inclusion of plasmonic structures will permit the realization of ultracompact active optical components such as optical transistors, photodetectors and modulators. However most of these mechanisms require conditions such as the application of heat or large external voltages that are often not practical when a device on chip needs to be fabricated for practical applications. Here, we propose a novel class of tunable metasurfaces in which the modulation is based on ionic conductance, permitting millivolt bias requirements apart from being applicable across plasmonic and dielectric metasurfaces.

Furthermore, if the scope of using plasmonic elements in future nanophotonic chips is to be increased, the problem of energy efficiency needs to be addressed. Not being able to reduce the inherent material losses, reducing the physical dimension of a device active volume will partially help alleviate this problem, however the need for attojoule level modulation is becoming a necessary requisite as well. It is known that the power consumption in a standard CMOS device goes as $P = CV^2 f$, where C is the capacitance of the device, V is the voltage bias applied and f is the clock frequency. Hence an ideal nanophotonic CMOS device will be significantly aided by displaying a low capacitance as well as very low operating voltages. Recent demonstrations of low energy optical modulators include a plasMOStor as a metal-oxide-Si field effect plasmonic modulator,[9] a femtojoule electro-optic modulation using a silicon-organic hybrid device,[10] atomic scale plasmonic switches,[11] and memristor based effects in plasmonic waveguides.[12] An interesting mechanism that some of the examples take advantage of, is that of ion conductance reminiscent of memristors.

Ion conductance arises in material systems that permit the movement of ions of a certain species across a host medium, especially under the application of an external bias. While permitting the movement of ions, such media prevent the movement of electrons, thus allowing for insulating behaviour despite the movement of ions. When the ions move, they cause a significant change in the resistance of the medium, thus giving rise to the mechanism of resistive switching. Resistive switching denotes a reversible phenomenon in two terminal elements, which change their resistance reversibly, often in a non-volatile manner, upon electrical stimuli. When the resistance states are non-volatile in nature, it is as if the stimulus affects an internal state variable of the element, which controls the resistance. This is akin to the resistance being *memorized* by the element, therefore comparing this mechanism to phenomena first proposed in the so-called *memristive elements*. Such a switching of resistance can be brought about by a wide variety of different phenomena including but not restricted to nanomechanical phenomena,[13] magnetoresistive effects such as spin-transfer torque (STT),[14] electrical effects such as leakage current through flash gate stacks in which trapping/de-trapping phenomena occur,[15] phase change between amorphous and crystalline phases,[16] and nanoionic redox phenomena or electrochemical metallization.[17] The particular mechanism which is utilized in this present work is based on the nanoionic redox phenomenon/electrochemical metallization. A typical system involves a simple metal-insulator-metal configuration. One of the metals is an active electrode that is oxidizable (soluble), such as silver or copper, while the counter electrode is inert (insoluble). An oxide, chalcogenide, or halide material in between these two electrodes serves to transport the metal cations. When a positive voltage is applied to the oxidizable electrode, its constituent metal starts to dissolve and results in the deposition of a metallic filament at the opposite inert electrode. In the extreme case, this metallic filament ultimately bridges the relatively insulating ion conductor and causes a very large change in the resistance of the device. Upon reversing the bias, the filament starts to dissolve and the system tends back towards its original state. Commonly used active electrodes include those with mobile ions, such as silver ions in silver halides and chalcogenides – AgI, $Ag_2S$, $Ag_2Se$, and $Ag_2Te$. It is well known that although the low-temperature crystalline phases of such materials are less conductive with often mixed ionic and electronic contributions, their high-temperature polymorphs are excellent conductors. However, what is also known is that the temperature at which they show excellent ionic conductance can be dramatically brought down to room temperature or lower simply by scaling down the size of the silver nanoparticles involved.[18] As for the insulator sandwiched in between the two electrodes, it is well known that the movement of ions in a medium is strongly influenced by interstitial channels in certain directions in crystalline materials, long-range disorder in amorphous, nanoscopically porous materials as well as the presence of defects including in conventional oxides such as alumina ($Al_2O_3$), silica ($SiO_2$) and titania ($TiO_2$).[19-22] Alumina deposited using atomic layer deposition has been shown to be conductive to specific sets of ions.[19] Finally, the counter electrode has in particular no other role to play, other than permitting the application of a bias.

Such a resistive switching based phenomenon has been attempted to be integrated with plasmonic systems, allowing the authors to explore switching in waveguides[12] as well as make atomic scale switches.[11] With the general mechanism of memristive switching requiring only a few 100 mV of biasing and continuing to go lower,[23] incorporating this mechanism in nanophotonics can perhaps help significantly lower the power consumption in optical chips. In this work, we experimentally demonstrate the integration of such a resistive switching mechanism in a plasmonic metasurface, permitting even lower operating voltages as small as 5 mV.

Our device consists of 80 nm of silver evaporated using electron beam evaporation with a 1 nm chromium adhesion layer onto a quartz substrate (SPI Inc.). Focused ion beam lithography is carried out on the deposited metal film to create a metasurface of inverse dolmen structures. The typical dimensions of these features involve the two lower parallel rods of lengths 150 nm and width 100 nm separated by a 30 nm gap. Both of these are separated by another 30 nm gap to the top rod of length 230 nm and width 100 nm. An array of 200 μm x 200 μm is milled onto the film and a subsequent 5 nm atomic layer deposition of $Al_2O_3$ is undertaken. Thereafter, 110 nm of indium tin oxide (ITO) is sputtered using RF magnetron sputtering (Ar+$O_2$ flow rate of 0.5 sccm) (see Fig. 1). All the deposition steps are undertaken with appropriate face masks made of stainless steel, at every step to permit an easy bias application configuration in the final device. Two sets of samples were made, one set was milled until the chromium adhesion layer was reached, and thereby retaining a thin layer of chromium still at the bottom of the milled inverse dolmen legs, while the other set was milled through till the quartz substrate was reached. We will refer to these two sets of samples as *with* and *without chromium* respectively in future references.

Materials characterization of the indium tin oxide was undertaken to determine the carrier concentration and permittivity of the deposited film, using the Van der Pauw Hall measurement technique (instrument make) and spectroscopic ellipsometry (VWASE, J. A. Woollam) respectively. The resulting carrier concentration was $0.5 \times 10^{20}/cm^3$ and the obtained permittivity is shown in the supporting information. The 5 nm ALD of alumina was repeatedly tested for its insulating properties and it was seen that most of the samples were displaying very good insulation properties (for sample curve see SI). Those samples that did not exhibit a good insulating layer were discarded. A transmission electron microscopy (TEM) image of an identical deposition on a silicon substrate was seen to show a uniform layer of roughly 1.2 nm of chromium as the adhesion layer (see SI for images) and hence did not create disconnected islands. Spectroscopic ellipsometry was conducted on a dummy sample of silver deposited on silicon, however it was observed that there was a temporal evolution of the permittivity. This was verified by undertaking spectroscopic ellipsometry every day for a week (see SI). It was observed that the trend in the permittivity would predict a slow red-shift of the intended plasmon resonance in our structures. Furthermore, energy-dispersive X-ray spectroscopy (EDX) was undertaken on the samples and it was seen that there was a finite amount of Sulphur present throughout, wherever silver was present (see SI). It is believed, that this was due to the general contamination of silver (tarnishing), when exposed to air. With various forms of materials characterization undertaken on our samples, optical measurements were undertaken to observe reflection, transmission and absorption.

All optical measurements were undertaken at normal incidence with incident polarization along the two parallel bottom legs (see Fig.). A microscope objective 5 X (Olympus, add NA) was used to focus down the incident light from a supercontinuum laser light source (Fianium) to a spot of size roughly xxx micron and the signal measured using a silicon detector (make specifications). Measurements were taken for a wavelength range from 450 nm to 850 nm. For the bias dependent measurements, a voltage was applied using a Keithley Source meter (specify make), and a bias applied between the top layer of ITO and the bottom layer of silver at appropriate locations far from the actual device area so as not to damage it. This was permitted by the design of the masks that created a pad large enough to be biased externally using either contact probes or conducting wires/tape.

The results of the optical measurements are shown in Fig.2. These results involve applying a positive bias to the silver electrode and a negative bias to the ITO electrode. This has two consequences. Firstly, this permits the movement of silver ions through the alumina, as was described in the earlier part of the manuscript and secondly, it creates a charge accumulation or active layer at the ITO-alumina interface. One can clearly see from Fig.2a, that upon the application of a bias of 5 mV (in 10 steps, leading to each incremental graph representing a bias of 0.5 mV), the reflection exhibits a broadband change in amplitude by nearly 25%. There is however no noticeable wavelength shift observed. Similarly, the measured transmission shows a decrease for the same set of bias voltages. And the consequent change in absorption shows an increase or decrease depending on the wavelength region looked at. The complexity of the curve can be attributed to the inherent complexity of the inverse dolmen structure itself. Also shown in the same figure are the full-wave electromagnetic numerical simulation results using finite difference time domain. It can be seen that the experimental plot is red-shifted with respect to the numerical results, by an amount which can be attributed to the temporal degradation of silver itself, as was discussed earlier in the manuscript. Another reason for this mismatch could be the non-uniformity of the milled structures across the focused spot. To qualitatively probe the phenomenon of resistive switching in our system, we conceptualize the silver ions as having moved into the alumina and possibly even the ITO, creating an effective medium in the region, whose complex permittivity depends on the fill-fraction of the silver ions in the 'host' medium. The presence of the silver ions at the ITO alumina interface also contributes to an increase in the decay parameter (decrease in the scattering time) of the ITO itself, which can be modeled by changing the 'gamma' parameter used in the Drude part to model the ITO. These effects are used to create an effective medium in the alumina region using the Bruggeman model of the effective medium theory. An increase in bias is modeled as an increase in the fill-fraction of the silver in the alumina, leading to a change in the complex permittivity. However at the same time, the gamma parameter of the inherent ITO is also changed (details in SI). As can be seen in Fig.2, probably the actual mechanism occurring in this system is a complex combination of the two above mentioned phenomena. The competition of these different factors determines the eventual trend observed experimentally. A complete thorough description of the exact modeling in our system is beyond the scope of this present work and will be undertaken in future work. However, what can be seen is that just the modification of the complex permittivity in the active region of ITO is not sufficient to explain our observed phenomenon, for with these small voltages, the index change is insufficient to observe what we experimentally obtain. Secondly, the modeling also hints that there are several competing factors contributing to our final experimental results. Nevertheless we can qualitatively see that these may be potential explanations for what we obtain.

A second interesting observation in our systems involves the distinction between samples with and without chromium. The obtained experimental reflection for the same polarization and set of applied voltages for the two sets of samples in seen in Fig.3. It is surprising to see how large an impact the chromium has on the observed phenomenon. As mentioned earlier, it is known that long-rage disorder as well as defects and stress in the insulating oxide can dramatically increase ionic conductance.[19] In particular, the presence of defects and stress on the behaviour of the ionic conductance of alumina (deposited using ALD) deposited onto chromium is known to (a) increase the oxygen vacancies as well as (b) reduce the surface resistance at the interface.[19] Apart from ionic conductance, the same resistive switching mechanism is also known to occur in systems exhibiting oxygen vacancies, such as in the $TiO_2$ based memristive systems.[24] Secondly, the reduction in surface resistance might be permitting quicker and easier movement of the silver ions in to the alumina layer. Therefore both these observed effects of the ionic conductance properties of alumina in the presence of chromium and chromium oxide, hint at the reason why such a large difference is seen in the two sets of samples. Therefore, it is imagined that the chromium is actually making the ionic conductance of the alumina much better and thereby permitting us to see this optical effect at such small voltage biases.

Upon the increase of the external bias, it is seen that the reflection increases until it reaches a maximum saturation value of 90% upon the application of 60 mV. Thereafter, there is a steep reduction in the reflection amplitude, tailing down at 12% beyond which there is a dramatic increase in the leakage current. This gives an amazing dynamic range of 78% with the application of just 100 mV before the breakdown occurs. It is believed that with this small 5 nm alumina layer and under these fabrication conditions, the filament formation is complete with as little as 100 mV, leading to a shorting of the insulating layer and increasing absorption and decreasing reflection significantly.

The metasurface was tested for its modulation speed by measuring the modulation of the reflection amplitude at normal incidence at a peak reflectance (lambda = xxx nm), with polarization along the bottom legs. It was seen that the measured reflection faithfully reproduced the modulation of the input bias up to 650 Hz. Since the modulation here is induced by the movement of ions, it is expected that it will cap out at a few kilohertz. However, since the same resistive switching mechanism can also occur in the form of other more mobile configurations such as oxygen vacancy defects, it is expected that modulation can go up to the GHz regime in other such similar systems.

In conclusion, in this letter we have experimentally demonstrated the first of its kind ultralow power consumption in an optical modulator exhibiting up to 78% modulation in reflection amplitude with a bias of just 0.1 V (and up to 25% with a bias of just 5 mV). Our system exploits the mechanism of resistive switching in metal-insulator-metal systems and improves upon previous work by pushing down the required bias voltage required for modulation by incorporating additional external defects that allows for higher ionic conductance. A complete understanding of such a mechanism can have a significant impact on the design of future attojoule optical modulators, highly sensitive optical readout of surface phenomena, optical display technology and even non-volatile optical memories.

AUTHOR INFORMATION

**Corresponding Author**

*kthyagar@caltech.edu, haa@caltech.edu

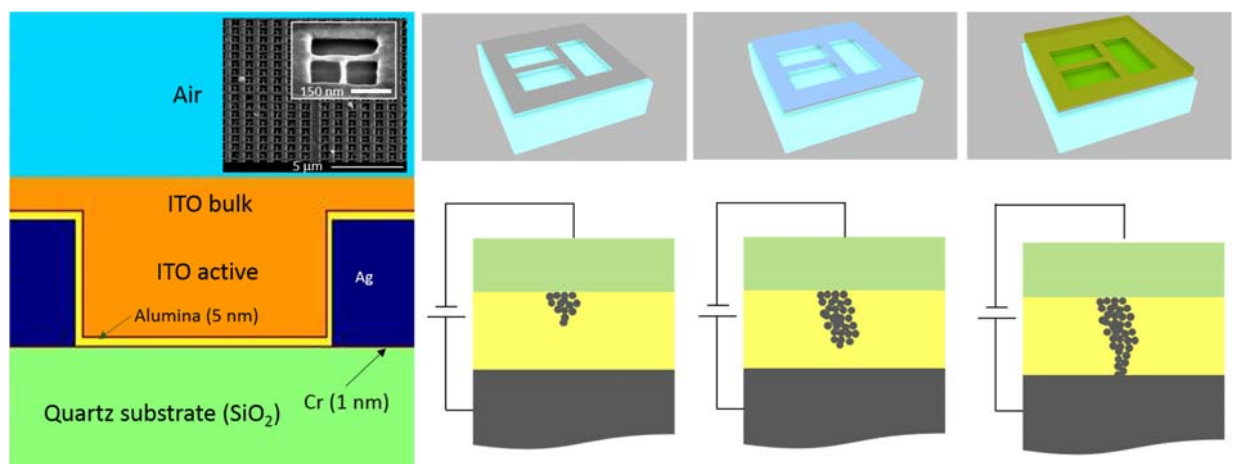

**Figure 1.**

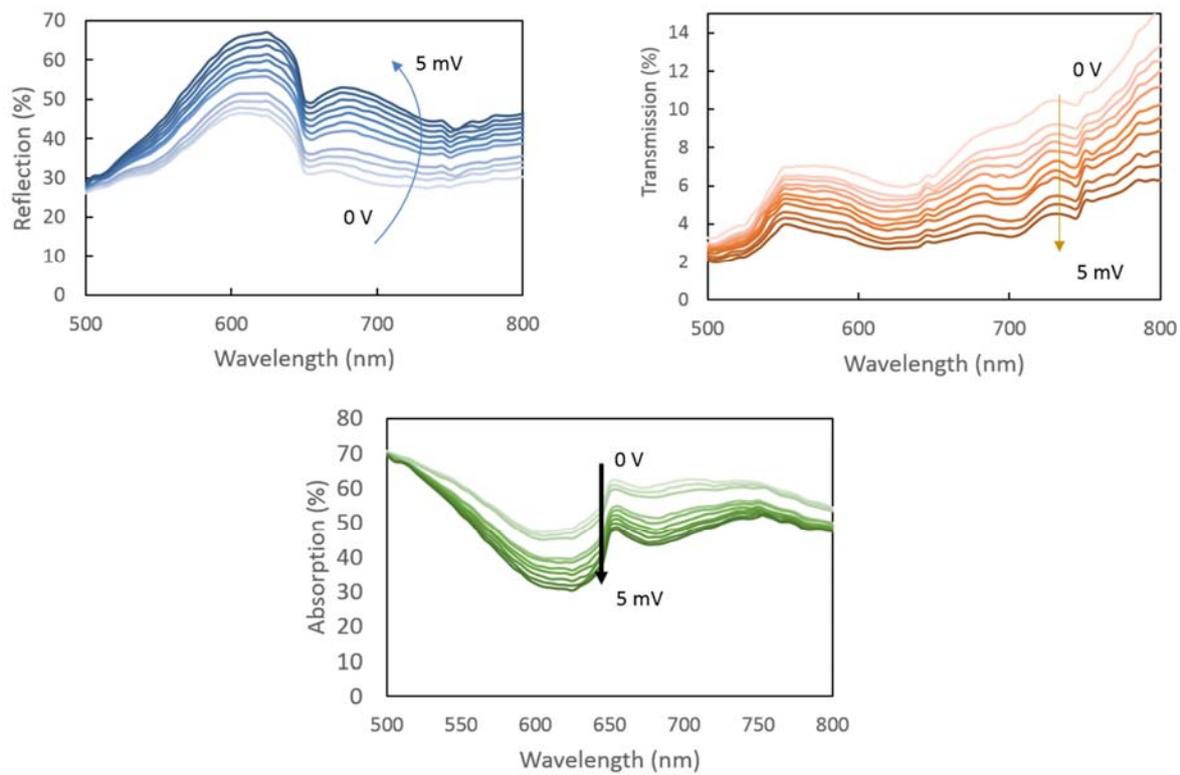

**Figure 2.**

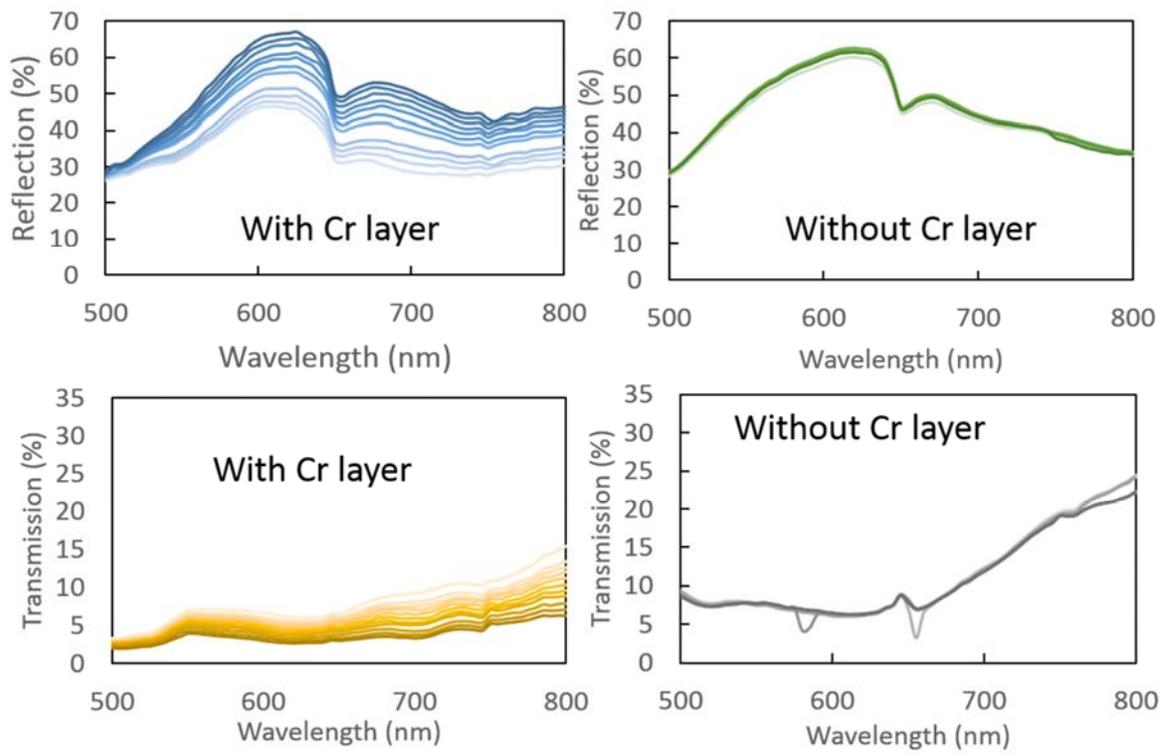

**Figure 3.**

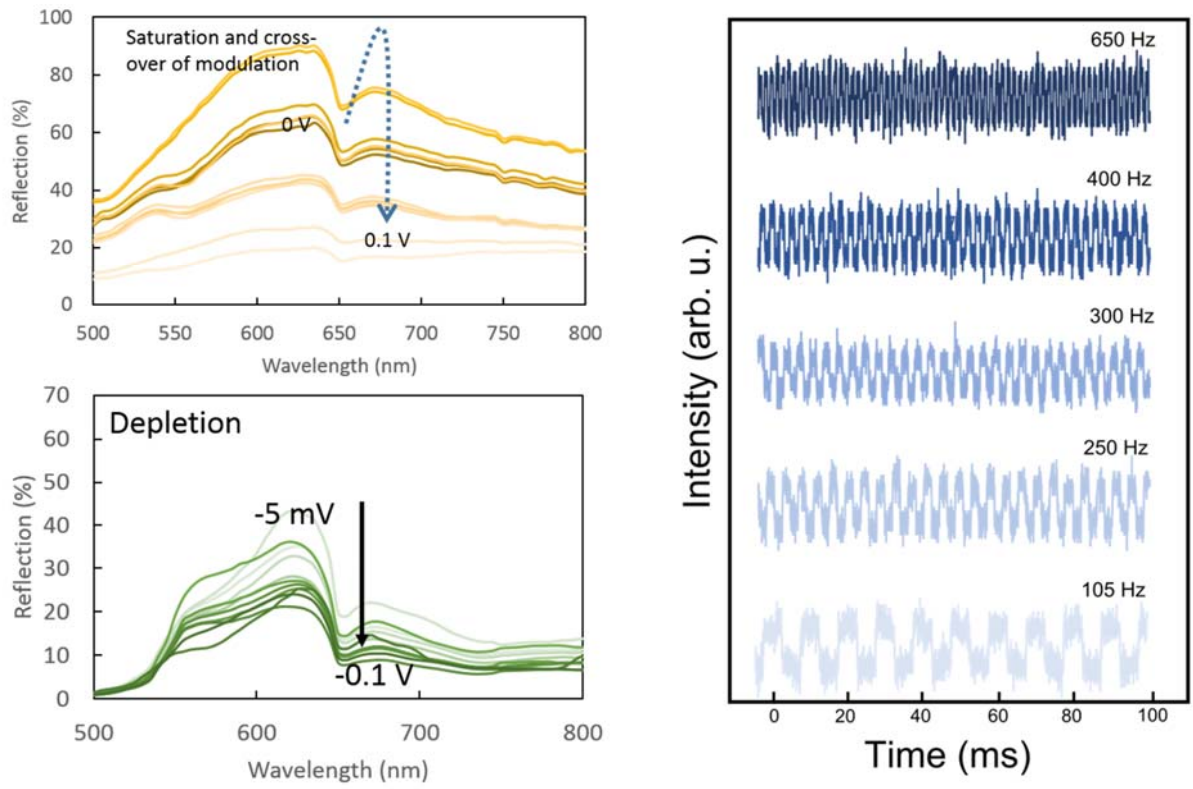

**Figure 4.**